\documentclass{article}
\usepackage{aaspp4}

\newcommand{\tsco}{$\tau$~Sco} 
\newcommand{\asca}{{\em ASCA}}
\newcommand{\rosat}{{\em ROSAT}}
\newcommand{\e}[1]{10^{#1}}

\newcommand{\mdot}{\.{M}}

\newcommand{\kms}{km~s$^{-1}$}

\newcommand{\Delv}{$\Delta \nolinebreak v$} 
 
\newcommand{\msun}{M$_{\odot}$}
\newcommand{\Msunyr}{{M$_{\odot}$ yr$^{-1}$}}

\newcommand{\osix}{O$\,${\small VI}}
\newcommand{\nfive}{N$\,${\small V}}

\newcommand{\teff}{$T_{\rm eff}$}
\newcommand{\vsini}{$v \sin i$}

\newcommand{\Mult}{${\cal M}(t)$}
\newcommand{\etal}{{et~al.}}

\begin{document}

\title{Stagnation and Infall of Dense Clumps in the Stellar Wind of
$\tau$~Scorpii}

\author{J. Christopher Howk\altaffilmark{1}, Joseph P. Cassinelli}

\affil{University of Wisconsin-Madison\\Department of Astronomy}

\author{Jon E. Bjorkman}
\affil{University of Toledo\\Dept. of Physics \& Astronomy}

\and

\author{Henny J.G.L.M. Lamers}
\affil{University of Utrecht\\Astronomical Institute}

\altaffiltext{1}{Current Address: The Johns Hopkins University, Dept. of
Physics and Astronomy, Baltimore, MD.  21218;\\ e-mail:~howk@pha.jhu.edu} 

\authoremail{howk@pha.jhu.edu}

\begin{abstract}

Observations of the B0.2 V star $\tau$ Scorpii have revealed unusual
stellar wind characteristics: red-shifted absorption in the
far-ultraviolet \osix\ resonance doublet up to $\sim+250$ \kms, and
extremely hard X-ray emission implying gas at temperatures in excess
of $\e{7}$ K.  We describe a phenomenological model to explain these
properties.  We assume the wind of \tsco\ consists of two components:
ambient gas in which denser clumps are embedded.  The clumps are
optically thick in the UV resonance lines primarily responsible for
accelerating the ambient wind. The reduced acceleration causes the
clumps to slow and even infall, all the while being confined by the
ram pressure of the outflowing ambient wind.  We calculate detailed
trajectories of the clumps in the ambient stellar wind, accounting for
a line radiation driving force and the momentum deposited by the
ambient wind in the form of drag.  We show these clumps will fall back
towards the star with velocities of several hundred \kms\ for a broad
range of initial conditions. The velocities of the clumps relative to
the ambient stellar wind can approach 2000 \kms, producing X-ray
emitting plasmas with temperatures in excess of $(1-6)\times \e{7}$ K
in bow shocks at their leading edge.  The infalling material explains
the peculiar red-shifted absorption wings seen in the \osix\ doublet.
Of order $10^3$ clumps with individual masses $m_c
\sim 10^{19} - 10^{20}$ g are needed to explain the observed X-ray
luminosity and also to explain the strength of the \osix\ absorption
lines.  These values correspond to a mass loss rate in clumps of
$\mbox{\.{M}}_c \sim \e{-9}$ to $\e{-8}$ \Msunyr, a small fraction of
the total mass loss rate ($\mbox{\.{M}} \sim 3 \times \e{-8}$
\Msunyr).  We discuss the position of \tsco\ in the HR diagram,
concluding that \tsco\ is in a crucial position on the main sequence.
Hotter stars near the spectral type of \tsco\ have too powerful winds
for clumps to fall back to the stars, and cooler stars have too low
mass loss rates to produce observable effects.  The model developed
here can be generally applied to line-driven outflows with clumps or
density irregularities.

\end{abstract}


\section{INTRODUCTION}

The star $\tau$ Scorpii (HD 149438) is the MK standard for the B0 V
spectral type, although Walborn (1995, private communication) has
classified it as B0.2 V on the basis of its stellar wind lines.  The
star has served as a benchmark in studies of radiatively driven winds
of hot stars.  This is because it lies in a pivotal region of the HR
diagram where stellar winds are making a transition from the massive
and fast winds of the O stars to the winds of B stars that can just
marginally be driven by line forces. 

In the Lamers \& Rogerson (1978) study of {\em Copernicus} satellite
spectra of early-type stars, \tsco\ was the latest spectral type to
show the superionization stage of \osix.  Stars with spectral types
close to that of \tsco\ are also the latest to show the X-ray to
bolometric luminosity relation $L_x /L_{bol}=10^{-7}$ that holds
throughout the main sequence O spectral region. For stars of spectral
type B1.5 V and later, this ratio of luminosities decreases by more
than an order of magnitude, (Cassinelli \etal\ 1994). It is generally
thought that the X-rays from hot stars arise from shocks embedded in
the stellar winds (Owocki, Castor, \& Rybicki 1988, hereafter OCR;
Cooper 1994), so the decrease in the X-ray luminosity may be an
additional consequence of the slower and less massive winds of near
main sequence B stars.  Massa, Savage, \& Cassinelli (1984) found from
a study of the UV lines that the wind properties of stars of spectral
types B0 V to B2 V are particularly sensitive to the stellar
composition, with winds ranging from non-detectable to quite strong
for stars with increasing metal abundances.  The stars near this
spectral range are also commonly seen as emission line Be stars, for
which the stellar rotation may produce wind compressed equatorial
disks as suggested by Bjorkman \& Cassinelli (1993) and Owocki,
Cranmer, \& Blondin (1994) or rotationally induced bistability as
suggested by Lamers \& Pauldrach (1991) and Lamers et al. (1999).

While \tsco\ is a standard star, its stellar wind is anything but
typical.  The two most unusual properties of its wind, and those that
we will focus on, are the peculiar red-shifted absorption seen in wind
lines of high ionization species such as \osix, which are not present
in the photosphere of this star, and its excessively-hard X-ray
spectrum.

The first of the peculiar wind properties of \tsco\ was discovered by
Lamers \& Rogerson (1978) in their analysis of {\em Copernicus}
observations.  They noted that the absorption produced by \osix\ and
\nfive\ extended {\it redward} of line center by as much as +250 \kms.
For stars with massive winds, the strong UV resonance lines have P
Cygni-type profiles, which are noted for the absorption shortward of
line center owing to scattering in the expanding wind.  The absorption
longward of the line center in \tsco\ is unusual.  Absorption redward
of line center could occur if the observed resonance line is also
present in the photosphere of the star.  However, the superionization
stages of \osix, \nfive, and \ion{P}{5} should not be present in the
photosphere of \tsco, which has an effective temperature of
$\sim31,000$ K.  Two ways to explain the redshifted component are
mentioned by Lamers and Rogerson (1978): turbulence in the flow near
the base of the wind, and infall of highly-ionized material. Both
phenomena could produce a positive velocity component in these ionic
lines. 

%The resonance lines of \osix\ and \nfive\ in \tsco\
%show wind absorption with radial velocities extending from $-1000$ to
%+250 \kms. 

The second peculiarity of \tsco\ is its very hard X-ray spectrum and
large X-ray luminosity.  In their \rosat\ survey of near-main sequence
B stars, Cassinelli \etal\ (1994) found that \tsco\ has a high X-ray
to bolometric luminosity ratio: $L_x / L_{\rm{bol}} \approx 10^{-6} $;
this is an order of magnitude larger than most O and early B stars.
Furthermore, they found that the X-ray spectrum has an anomalously
luminous hard component above 1 keV.  The hard X-rays from this wind
are unusual in that they require plasma temperatures in excess of
$\e{7}$ K.  Cohen, Cassinelli, \& Waldron (1997; hereafter CCW) have
more recently presented observations of \tsco\ obtained with the
\asca\ satellite.  These observations confirm the earlier \rosat\
results requiring a very high temperature component to explain the
observed X-ray spectrum.  The \asca\ observations were fit with a
multi-temperature equilibrium plasma model characterized by
temperatures of $T = (7, \ 12, \ \mathrm{and} \ 27) \times \e{6}$ K
with emission measures of $EM_X = (3.5, \ 0.8, \ \mathrm{and} \
>0.3)\times \e{54}$ cm$^{-3}$, respectively.

The very high temperature gas inferred from the X-ray spectrum of
\tsco\ is not explained in the standard picture of X-ray production in
early type stars (see discussion in CCW).  Numerical simulations of
radiative shocks formed in the unstable stellar winds of OB stars
suggest that these shocks are unable to produce temperatures much
greater than a few times 10$^6$ K (OCR; Cooper 1994).  The shock-jump
velocities in these simulations are limited to roughly half the local
stellar wind velocity.  The presence of very hot gas in the wind of
\tsco\ suggests that a different mechanism is responsible for at least
some of the X-ray emission from this star.  Though several groups have
attempted to model the radiatively-driven outflow from
\tsco, they were unable to reproduce the observed hardness of the
X-ray spectrum (e.g., Cooper 1994; MacFarlane \& Cassinelli 1989; Lucy
\& White 1980).

In this paper we explore a phenomenological model to explain both the
hard X-ray emission and the redshifted absorption of high ionization
lines. Our model supposes density enhancements, or clumps, form in the
smooth stellar wind and subsequently fall back towards the stellar
surface, creating bow shocks at their leading edges.  We investigate
the trajectories of such clumps in a stellar wind and find infall to
be achievable assuming the optical depths of these dense parcels of
material are large enough to choke off the radiative line driving
force.  Using our calculated trajectories, we estimate the temperature
of the shocked gas swept up by the clumps and comment on the ability
of this model to provide the high-temperature gas implied by the
observed X-ray emission (CCW; Cassinelli \etal\ 1994).  While the
numerical details presented here apply to the specific case of \tsco,
the model developed in this paper is applicable to most clumped,
radiatively-driven outflows.

The details and assumptions of our model are discussed in \S
\ref{sec:model}.  In \S \ref{sec:infall} we derive the equation of
motion for solitary clumps created within the wind and present the
results of numerical integrations of clump trajectories.  We describe
the temperature spectrum of shocked plasmas in our model and discuss
the observed X-ray emission in the context of this model in \S
\ref{sec:xrays}.  In \S \ref{sec:conclusions} we briefly discuss why
\tsco\ may be unusual among early-type stars, and we summarize our
work in \S \ref{sec:summary}.

Table \ref{table:stardata} lists the stellar parameters adopted
throughout this work for \tsco.  Empirical determinations of the mass
loss rate of \tsco\ from UV lines gives discrepant results of
$7\times10^{-9}$ (Lamers \& Rogerson 1978), $7\times10^{-8}$ (Gathier
et al. 1981) and $ 1\times10^{-9}$ \Msunyr\ (Hamann 1981). These
differences are due to differences in the adopted ionization fractions
of the observed UV resonance lines of \ion{C}{3}, \ion{C}{4},
\ion{N}{3}, \nfive, \osix, \ion{Si}{4}, and \ion{P}{5}.  The terminal
velocity of the wind of \tsco\ cannot be derived from the
observations, because the UV resonance lines are unsaturated.  Lamers
and Rogerson (1978) derived a lower limit of 2000 \kms.  Because of
these uncertainties we adopt the values predicted by the modified line
driven wind theory (Kudritzki \etal\ 1989) with the force multiplier
parameters taken from Abbott (1982): $k$=0.156, $\alpha$=0.609 and
$\delta$=0.12.  This gives a mass loss rate of $3.1\times10^{-8}$
\Msunyr\ and $v_\infty$ = 2400 \kms, in agreement with the range of
empirically derived parameters.

%It is noteworthy that \tsco\ has an unusually low observed
%rotational velocity.  The observed rotational velocity is \vsini $\ 
%\approx 20$ \kms\ (Uesugi and Fukuda, 1982), while the average for
%\sptype\ stars is $\langle v \sin i \rangle \approx 200$ \kms\ (Allen
%1973).  This implies that \tsco\ is either a slow rotator or is seen
%nearly pole-on.  Also, Waters, \etal\ (1993) have observed unusual
%infrared hydrogen emission lines associated with \tsco.  The lines of
%\Bra\ and \Brg\ are seen in emission, though there is no evidence for
%corresponding H$\alpha$ emission.  They speculate that \tsco\ may be a
%`hidden' Be star.  They argue that the observed hydrogen line
%intensities and shapes could be reproduced by a low density disk if
%the star is seen nearly pole-on (Waters, \etal\ 1993; Zaal, Waters \&
%Marlborough 1995).

%\begin{table}[ht]        

\section{THE MODEL}
\label{sec:model}

The winds of early type stars are driven by resonance line scattering
of radiation in the stellar envelope.  This radiative driving
mechanism is inherently unstable (Lucy and Solomon 1970; Lucy 1982;
OCR).  Small perturbations on an otherwise monotonic wind velocity
structure tend to grow and steepen into shocks. These instabilities
have been well studied using one-dimensional hydrodynamical models
(see, e.g., Cooper 1994, Feldmeier 1995), which show that the
resulting wind is a mixture of rarified hot material and colder, dense
regions.  These studies also suggest the shock velocity jumps have
maximal amplitudes of order one half the local wind velocity.  The
relationship between the velocity amplitude of a shock, \Delv, and the
temperature behind the shock, $T_s$, is given by
\begin{equation}
T_s=1.44 \times 10^{5} \left( \frac{\Delta v}{100 \ \mathrm{km/s}} 
	\right)^2 \ {\mathrm K}.
\label{eqn:tshock}
\end{equation}
In the case of most main sequence O stars and OB supergiants, the
shock jump velocities are indeed strong enough to produce gas at the
temperatures suggested by the observed X-ray emission ($T \approx
3\times \e{6}$ K; Chlebowski, Harnden, \& Sciortino 1989; Cassinelli
\etal\ 1994). The X-ray spectrum of \tsco, however, implies gas with
temperatures in excess of 10$^7$ K (CCW).  Numerical models of B star
winds cannot produce shocks with such high temperatures.  Velocity
jumps of $\sim300$ \kms\ are typical in numerical simulations (Cooper
1994), while velocity jumps of more than 700 to 800 \kms\ are required
to reproduce the hard X-ray emission from \tsco.

While the one-dimensional shock production problem has been well
studied, little attention has yet been given to the production of
shocks and the structure of the wind in two or three dimensions.  We
assume that shocks and density enhancements in the stellar wind are
relatively localized phenomena whose sizes are dictated by the local
sound speed and instabilities that may tend to break up large sheets,
such as the Rayleigh-Taylor instability.  An individual shock is
assumed to cover a small solid angle when viewed from the star.  The
one-dimensional simulations assume spherical symmetry; our assumption
is that the spherical shocks in these simulations would fragment
relatively easily.  Cassinelli and Swank (1983) have suggested such a
scenario to explain the lack of variability observed in their Einstein
Solid State Spectrometer observations of OB supergiants.  A stellar
wind containing an ensemble of $N$ shocks that continuously form and
dissipate can produce a non-detectable variability of order
$\sqrt{N}/N$.  The time variability of single spherical shocks in the
wind of \tsco\ is discussed by MacFarlane \& Cassinelli (1989).

\begin{figure}
\epsscale{0.75}
\plotone{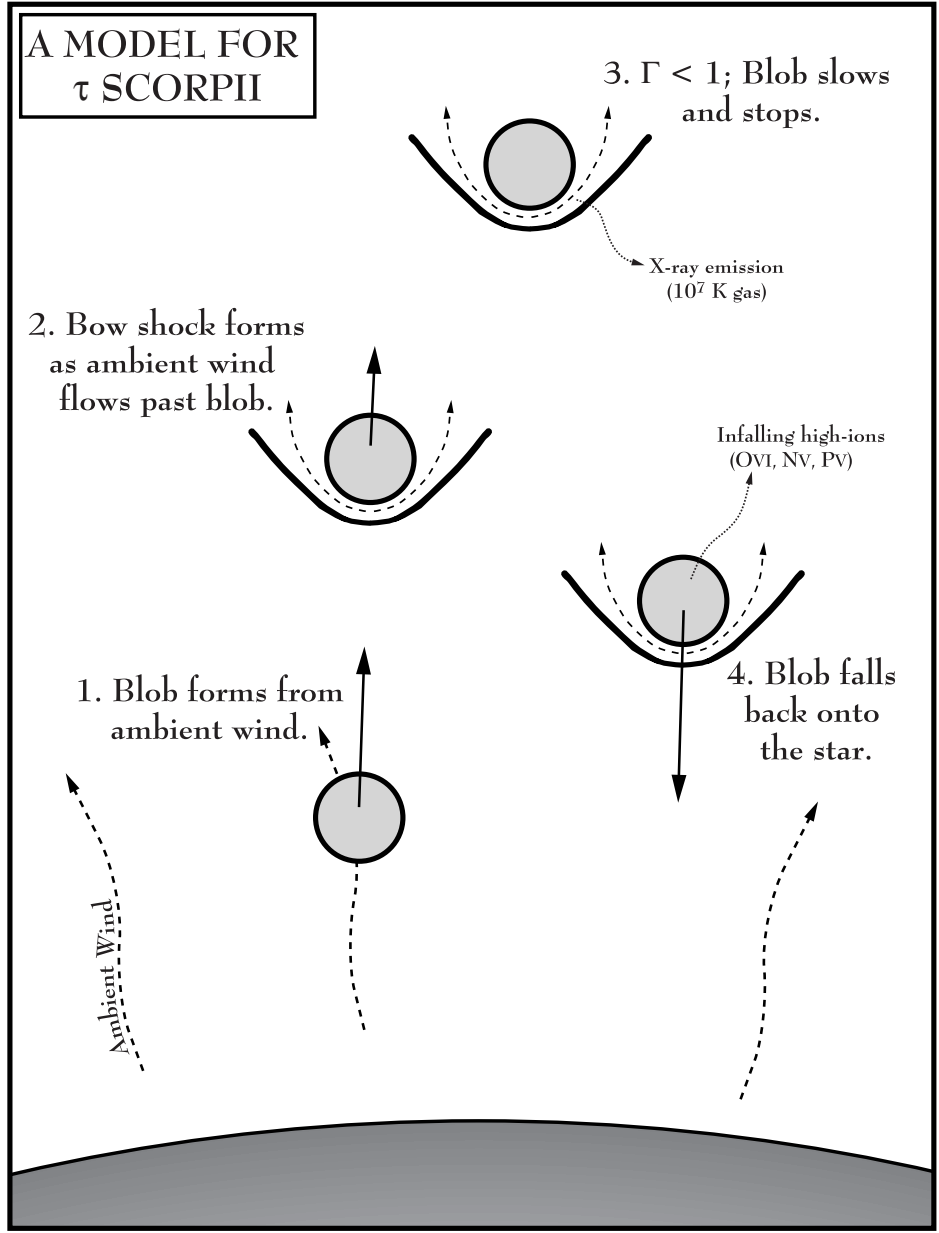}
\figcaption{Illustration of the production of X-rays and infalling
matter in the form of clumps in the stellar wind of \tsco.  The clumps
are formed from the ambient stellar wind.  Because the radiation force
is lower than that of gravity ($\Gamma_e + \Gamma_L < 1$), the clumps
slow, stop, and then fall back onto the star.  The hard X-rays are
formed in a bow shock where the wind collides with an infalling
clump. The X-rays produce the superionization seen in the UV spectra by
way of the Auger ionization mechanism. The wind is also assumed to
have the shocks typically asssociated with hot stars; these produce
the soft component of the X-ray spectrum of \tsco.
\label{fig:cartoon}}
\end{figure}

A schematic of our model is shown in Figure \ref{fig:cartoon}.  The
basic picture of the stellar wind that we adopt is one in which
discrete density enhancements, or clumps, are embedded in an otherwise
smooth stellar wind.  These dense clumps of material have shock
interfaces with the faster moving ambient wind.  The most extreme of
these clumps are dense enough to be optically thick to radiation in
the strongest resonance lines, though still optically thin to UV
continuum radiation.  In addition to their treatment of stellar winds
with organized velocity gradients, the theory of radiative
acceleration on material with no internal velocity gradient was
developed by Castor, Abbott, \& Klein (1975, hereafter CAK), and we
use their parameterization of the line-driving force.  The equations
governing this acceleration are given in \S
\ref{sec:infall}.  Assuming a CAK parameterization of the line driving
force, the net radiative acceleration on optically thick parcels of
material with no internal velocity gradient can be relatively small
(see \S \ref{sec:infall}).  Gravity dominates the motion of these
optically thick clumps, which fall inward, creating bow-shocks at
their leading edges as they plow supersonically through the outflowing
wind.  The velocity structure of the ambient wind is assumed to be
relatively unaffected by such clumps.  They do not shadow a
significant amount of the overlying wind, since they are optically
thin to continuum radiation and the line opacity is significantly
redshifted relative to the opacity driving the unperturbed wind.

As the clumps fall inward, they shock the out-flowing gas of the
ambient (inter-clump) wind. The temperature of gas the immediate
post-shock region is proportional to the square of the shock jump
velocity as in Equation (\ref{eqn:tshock}).  Since the relevant
velocity for the calculation of the shock temperatures in these bow
shocks is the {\em relative} velocity between the clumps and the
ambient wind, the temperatures needed to produce the hard X-ray
component observed by \rosat\ and \asca\ are readily obtained.  The
X-ray producing shocks adjacent to the leading edge of the infalling
clumps produce very highly-ionized material, thus accounting for the
red-shifted high ions seen in absorption (\osix, \nfive, etc.).

A related model was considered by Lucy \& White (1980, hereafter LW),
in the first paper attempting to explain the source of hot star X-rays
as shock-heated gas. They suggested that clumps are radiatively driven
outward while the rest of the wind is pushed along by the drag of
these objects. In our model the clumps move outward (or inward) with a
velocity less than that of the ambient wind, so the shocks occur on
the inner face of the clump. This placement of the shocks is more
consistent with the results of OCR, who find that shocks tend to be
inward facing with low-density, high-speed wind material being driven
into slower and denser shocked regions. However, we retain the clump
aspect of the LW model.

Let us assume the clumps are self-contained parcels of gas confined by
the ram pressure of the wind through which they fall.  Following LW,
we assume a balance of dynamical pressures and write the average
density of a clump, $\langle \rho_c \rangle$, falling through the wind
as
\begin{equation}
        \langle \rho_c \rangle = \rho_w (r) 
                \frac{ ( a_w ^2 + C_{\rho} \omega^2 )}{a_c ^2}. 
        \label{eqn:dens}
\end{equation}
In this expression $\rho_w$ is the wind density, a function of the
distance $r$ from the star, $a_w$ and $a_c$ are the thermal velocities
in the wind and clump, respectively, and $\omega$ is the relative
velocity between the wind and the clump.  We have ignored the effects
of turbulence within the clumps but have included the constant
$C_{\rho} = 0.29$ as a scale factor to account for the range of angles
at which the wind hits the bow-shock (see LW).  We can see from this
that once a clump begins to decelerate, thereby increasing $\omega$,
the oncoming wind confines it more effectively, making it denser as it
falls towards the star.  This further decreases the ability of the
radiation field to drive the clump outwards.  If the mass of material
accumulated into a clump is specified, then the volume can be readily
obtained from the density derived using Equation (\ref{eqn:dens}).

This model has the desirable attribute of producing both the
red-shifted high-ion absorption and the anomolously hard X-rays.
Several independent lines of evidence suggest that clumping can be
important in the winds of Wolf-Rayet (e.g., L\'{e}pine, Moffat, \&
Henriksen 1996) and OB supergiant winds (e.g., Sako \etal\ 1999;
Eversberg, L\'{e}pine, \& Moffat 1998).  It is within reason that near
main sequence early type stars may also contain localized regions of
significantly higher than average densities.  The exact mechanism for
the formation of these dense clumps is not rigorously approached in
this work, but it is worthwhile to explore the consequences of such a
model before a more thorough hydrodynamical study is undertaken.

\section{INFALLING MATERIAL}
\label{sec:infall}

The main driving force for winds from massive stars is the radiative
acceleration associated with atomic absorption line opacity.  The most
important of the driving lines are the ultraviolet resonance lines of
abundant ionic species.  Thus, the driving force is not only dependent
upon the elemental abundances in a wind, but also upon the wind's
ionization balance (MacFarlane, Cohen, \& Wang 1994; Abbott 1982).
The equation of motion describing the trajectory of a parcel of gas
with respect to the star is
\begin{equation}
{\frac{d v_c}{d t}} 
        = -\frac{GM_*}{r^2} (1-\Gamma_e - \Gamma_L) + f_d, 
   \label{eqn:motion1}
\end{equation}
where $v_c$ is the velocity of the clump of material, $M_*$ the mass
of the star, $r$ the distance of a clump from its center, and $f_d$ is
the force per unit mass the ambient wind exerts upon the clump in the
form of a ``drag."  The terms $\Gamma_e$ and $\Gamma_L$ are the ratios
of the radiative to gravitational acceleration for electron scattering
and resonance line scattering, respectively.  The drag force, $f_d$,
is zero if there is no relative motion between a clump and the ambient
wind.  In this treatment we have not allowed for a thermal pressure
gradient in the wind, which is negligible in the supersonic portion of
such an outflow.

For the ambient stellar wind, $f_d=0$.  The requirement for the
material to be accelerated away from the star is $d v / d t > 0$,
implying $\Gamma_e + \Gamma_L > 1$, i.e., that the radiative
acceleration is greater than the gravitational acceleration.  A dense
clump of material formed in the wind that has $\Gamma_e + \Gamma_L <1$
may decelerate and possibly fall back to the surface of the star,
plowing its way through the out-flowing wind as it falls.  However,
even for clumps with $\Gamma_e + \Gamma_L<1$, the drag force, $f_d$,
may be great enough to push the clump outwards in the stellar wind.
To determine if infall is physically reasonable, we shall develop more
fully each of the terms in Equation (\ref{eqn:motion1}).

\subsection{Radiative Acceleration}

The radiative driving force on a parcel of wind material is considered
to have two components: radiative driving via electron scattering and
resonance line opacities.  The radiative acceleration due to electron
scattering of the stellar continuum is written
\begin{equation}
        g_{e} = \frac{\sigma_e F}{c};
\end{equation}
here $\sigma_e$ is the electron scattering opacity, which is $\sigma_e
= \sigma_{Th}/ \mu_e m_H \approx 0.34$ cm$^2$~g$^{-1}$ for \tsco, $F$
is the frequency-integrated stellar flux at the location of the
material, and $c$ is the speed of light.

A proper treatment of the radiative driving of stellar winds owing to
absorption line opacity has been given by CAK (see also Abbott 1982).
Their theory paramaterizes the line driving force on a wind having an
organized velocity gradient $dv /dr$.  In CAK theory, the radiative
acceleration due to line scattering, $g_L$, is written
\begin{equation}
        g_L =  \frac{\sigma_e F}{c} {\cal M} (t) = 
	g_{e}^{\rm ref} {\cal M} (t).
\label{eqn:gline}
\end{equation}
where $g_e^{\rm ref}$ is for a reference electron scattering opacity
of 0.325 cm$^2$ g$^{-1}$.  So $g_L$ is simply the reference electron
scattering acceleration modified by a multiplicative factor.  This
factor, \Mult, the so-called ``force multiplier,'' is a function of
$t$, an optical depth parameter that is independent of line strength.
The force multiplier is defined as
\begin{equation}
{\cal M} (t) \equiv kt^{-\alpha} \left( \frac{N_e}{W} \right)^{\delta},
\label{eqn:multdef}
\end{equation}
where $N_e$ is the electron density in units of $10^{11}$ cm$^{-3}$,
$W$ is the dilution factor and $k, \ \alpha$ and $\delta$ are the CAK
parameters.  These parameters have been determined from studies
employing extensive atomic line lists and depend upon the temperature
of the absorbing material (see, e.g., Abbott 1982).  However, the
driving force for the density enhancements we will consider here is
slightly different from the typical CAK driving force for the ambient
wind.

In the CAK treatment of expanding atmospheres or winds, $t$ is the
electron scattering optical depth over the distance in which the wind
velocity increases by the thermal velocity. Hence $t$ is inversely
proportional to the local velocity gradient for outflows with
monotonically increasing velocity laws (CAK).  In the case of parcels
of gas with no organized velocity gradient, such as those being
considered here, the optical depth parameter $t$ is the integral of
the electron scattering opacity over the path length $l$ through the
region or clump [Equation (5) of CAK]:
\begin{equation}
t \equiv \sigma_e \int_l \rho dr.
        \label{eqn:tdef2}
\end{equation}  
Approximating the integral as $\langle \rho_c \rangle V_c^{1/3}$,
where $V_c \equiv m_c/\rho_c$ is the volume of the clump, Equation
(\ref{eqn:tdef2}) may be rewritten, using Equation (\ref{eqn:dens}),
as
\begin{equation}
     t \approx \sigma_e m_c^{1/3} \rho_w(r) ^{2/3} 
	   \left( 1 + \frac{C_\rho \omega^2}{a_c^2} \right) ^{2/3}, 
\label{eqn:t2}
\end{equation}
assuming $a_c \approx a_w$.  This latter assumption has the desirable
consequence that the density of a clump will approach that of the
ambient wind as the relative velocity, $\omega$, approaches zero.

The ratio of the radiative to gravitational accelerations in CAK
formalism is written
\begin{equation}
  \Gamma_e + \Gamma_L = \left[ \frac{\sigma_e L}{4 \pi c G M_*} \right] 
	\left[ 1 + {\cal M}(t) \right],
\label{eqn:gamma}
\end{equation}
given $L$, the luminosity of the star.  We will assume the values of
the CAK parameters tabulated by Abbott (1982) for stellar winds at
30,000 K ($k=0.156, \ \alpha=0.609$ and $\delta=0.12$) are appropriate
for use in the clumps considered here.  These are the same as those
used for estimating the mass loss rate.  Given the high densities
predicted for the clumps and the presence of a hard X-ray source at
their leading edges, the ionization balance in the clumps could be
different than that of the ambient wind.  Thus while the CAK
parameters appropriate for the clump may be different than those used
here, we adopt these parameters in order to proceed.  In the end, the
choice of CAK parameters will likely not matter too much so long as
the optical depth through a clump is very high.  Equation
(\ref{eqn:gamma}) gives an upper limit to the radiative acceleration
because it assumes that the flux from the star reaching a parcel of
gas is unattenuated by the layers closer to the star.  For a clump
moving outwards at lower velocities than the ambient wind, the photons
producing the line radiation pressure on the dense clump may already
have been absorbed or scattered by gas lower in the wind with the same
outward velocity as the clump.  We will ignore this effect here, which
is expected to be small.

\subsection{Drag force}

The standard drag force per unit mass on a clump of area $A_c$ moving
through an ambient medium with a relative velocity $\omega$ is given
by
\begin{equation}
f_d = \frac{1}{2} C_D \rho_w (r) \omega^2 \left( \frac{A_c}{m_c} \right).
\end{equation} 
In this treatment $m_c$ is the mass of the clump, and $\rho_w(r)$ is
the density of the ambient medium (in this case the ambient stellar
wind).  The drag coefficient, $C_D$, dictates the efficiency of
momentum transfer to the clump, and the factor of $\onehalf$ is
conventionally included.  This expression describes the energy density
impinging on the clump integrated over the area of the clump.  For
most of this work we will assume $C_D = 2$, i.e., perfect transfer of
momentum from the ambient wind to the clump.  This is the most
stringent assumption for determining if infall is viable.

The cross section of the clump may be approximated $A_c \approx
V_c^{2/3}$, giving
\begin{equation}
f_d = \frac{\frac{1}{2} C_D \rho_w \omega^2}
	{ \left( m_c \langle \rho_c \rangle ^2 \right)^{1/3}}. 
\end{equation}
Using Equation (\ref{eqn:dens}), and assuming $a_c
\approx a_w$ and $\omega \gg a_c$, we rewrite this as
\begin{equation}
    f_d \approx \frac{1}{2} C_D \left( \frac{a_c^4 \, \rho_w
    \omega^2}{C_\rho ^2 \, m_c} \right) ^{1/3} .
\label{eqn:drag}
\end{equation}
The density of the wind at a distance $r$ from the surface of the star
is calculated assuming mass conservation:
\begin{equation}
	\rho_w (r) = \frac{\mbox{\.{M}}}{4 \pi r^2 v_w (r)},
\label{eqn:continuity}
\end{equation}
where \mdot\ is the mass loss rate (see Table \ref{table:stardata}).
The velocity of the wind $v_w(r)$ is calculated using a $\beta$
velocity law with $\beta_w = 0.8$ (Groenewegen \& Lamers 1989):
\begin{equation}
    v_w (r) = v_\infty \left( 1- \frac{R_*}{r} \right) ^{\beta_w}.
\label{eqn:windv}
\end{equation}
with the values of $v_\infty$ and $R_*$ given in Table
\ref{table:stardata}.

The magnitude of the drag force can be relatively small.  We estimate
its magnitude using Equations (\ref{eqn:drag}) and
(\ref{eqn:continuity}) with the data given in Table
\ref{table:stardata}:
\begin{equation}
f_d \approx 55 \ \mathrm{cm \ s^{-2}} \ 
	\left( \frac{r}{R_*} \right)^{-2/3} 
	\left( \frac{m_c}{\e{19} \, {\mathrm g}} \right)^{-1/3} 
		v_w(r)^{1/3},
\end{equation}
where we have assumed $\omega = v_w$, which in this equation is in
units \kms.  If we define $\Delta \equiv f_d/g$, where $g$ is the
force of gravity per unit mass, we can rewrite Equation
(\ref{eqn:motion1}) as $d v_c / d t = -g(1 - \Gamma_e - \Gamma_L -
\Delta)$.  We find
\begin{equation}
  \Delta \approx 6\times\e{-3} \ 
	\left( \frac{r}{R_*} \right)^{4/3} 
	\left( \frac{m_c}{\e{19} \, {\mathrm g}} \right)^{-1/3} 
		v_w(r)^{1/3},
\end{equation}
where $v_w$ is again in \kms.  We can see that very near the star, the
drag force is relatively small compared with the force of gravity
because the velocity between the clump and the ambient wind is smaller
than at larger radii.  This prediction is based on the assumption
$\omega \approx v_w$, but for clumps producing temperatures in excess
of $\e{7}$ K, our numerical calculations discussed in \S
\ref{subsec:traject} suggest the drag force is much more significant (relative
to gravity) at larger radii.  For clumps with masses $\log m_c \approx
20.0$ (in grams) the outward drag force is comparable to the force of
gravity at large distances ($r > 10 R_*$) from the stellar surface.
For lower mass clumps, however, the drag force can play an important
role in the dynamics at intermediate distances from the star ($r\sim5
R_*$).

\subsection{Numerical Results}
\label{subsec:traject}

We have derived the individual components of Equation
(\ref{eqn:motion1}), and can hence obtain trajectories in
position-velocity space for clumps in the wind of \tsco.  Equation
(\ref{eqn:motion1}) shows the dynamics of such clumps depend on the
relative velocity, $\omega$, between the clump and the ambient wind:
$\omega \equiv v_w(r) - v_c$, where $v_w(r)$ and $v_c$ are the wind
and clump velocities, respectively.  We have numerically integrated
the equation of motion for parcels of gas with various initial
conditions since it does not have a analytically tractable solution.
The initial conditions specified in our approach are the initial
distance $r_{\circ}$ from the center of the star, the initial relative
velocity $\omega_{\circ}$, and the mass $m_c$ of the clump (we assume
mass conservation).  By specifiying $\omega_{\circ}$ and $r_{\circ}$,
we are also then specifying the initial velocity of the clump, given
Equation (\ref{eqn:windv}).  We assume the value of $\omega_{\circ}$
is characteristic of the jump velocities typically seen in numerical
simulations, i.e., $\Delta v \la v_w /2$ (Cooper 1994).  The
dependence on $m_c$ comes in through Equations (\ref{eqn:t2}) and
(\ref{eqn:drag}), the former affecting \Mult\ in the radiative
acceleration.

\begin{figure}
\plotone{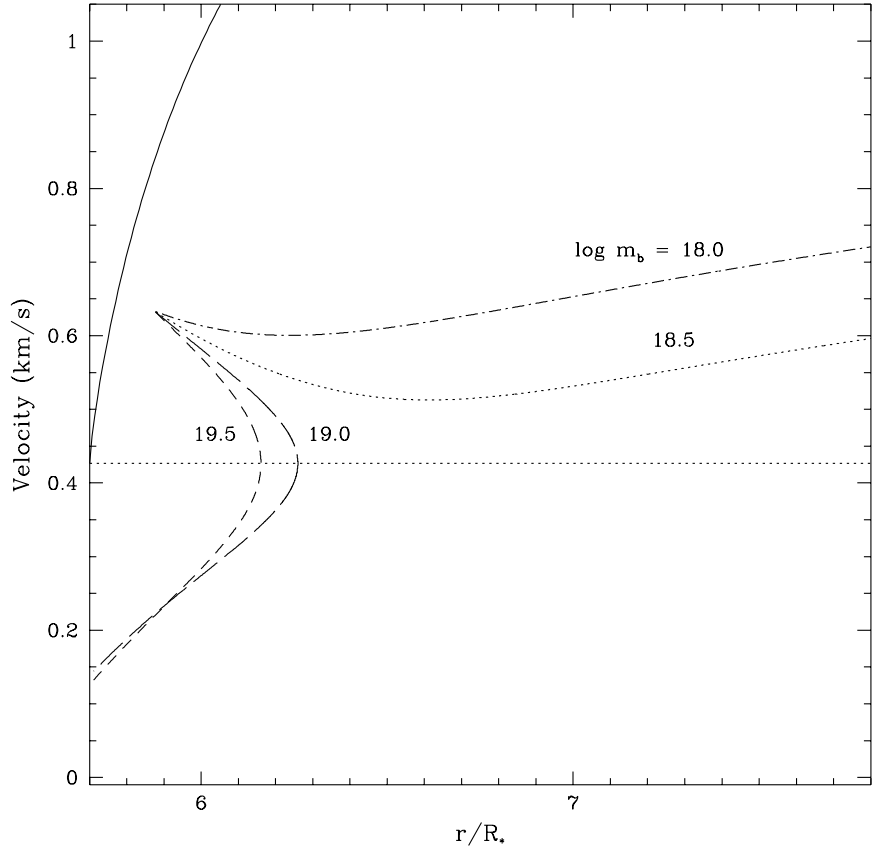}
\figcaption{Trajectories of clumps forming in the wind of $\tau$ Sco
having different initial conditions.  The velocity scale is relative
to the surface of the star, and the distance from the stellar surface
is given in units of stellar radii.  Also shown is the $\beta$
velocity law of the ambient stellar wind (solid line).  The starting
distance from the star and velocity relative to the ambient wind for
these clumps are $(r_o, \ \omega_o) = (1.25 \ R_*, \ 331 \ \mathrm{km \
s^{-1}})$.  The four different trajectories represent clumps formed
with masses $\log m_c = 18.0$, 18.5, 19.0, and 19.5 ($m_c$ in grams).
The masses are labelled for each trajectory.  Note the low-mass clumps
are pushed outwards through the wind, following {\em outflow}
trajectories.  The higher mass clumps are able to fall back onto the
star and follow {\em infall} trajectories.
\label{fig:trajectory}}
\end{figure}

Our numerical integrations of the equation of motion show two types of
clump trajectories: outflow and infall trajectories.  We term those
trajectories in which the clumps are driven from the star, though they
may initially decelerate, {\em outflow} trajectories.  Though these
clumps are driven outward, they always have velocities less than the
terminal velocity of the wind.  Those that decelerate and eventually
descend towards the star follow {\em infall} trajectories.  The latter
type is the most interesting, given the redshifted high ion absorption
line observations.  It is not clear, however, that outflow and infall
trajectories are mutually exclusive. There may exist a mix of
trajectories based upon a distribution of initial conditions for the
clumps.  Figure
\ref{fig:trajectory} displays the velocity of representative outflow
and infall trajectories as a function of distance from the star. Also
shown is the adopted $\beta$ velocity law of the wind.

\begin{figure}
\plotone{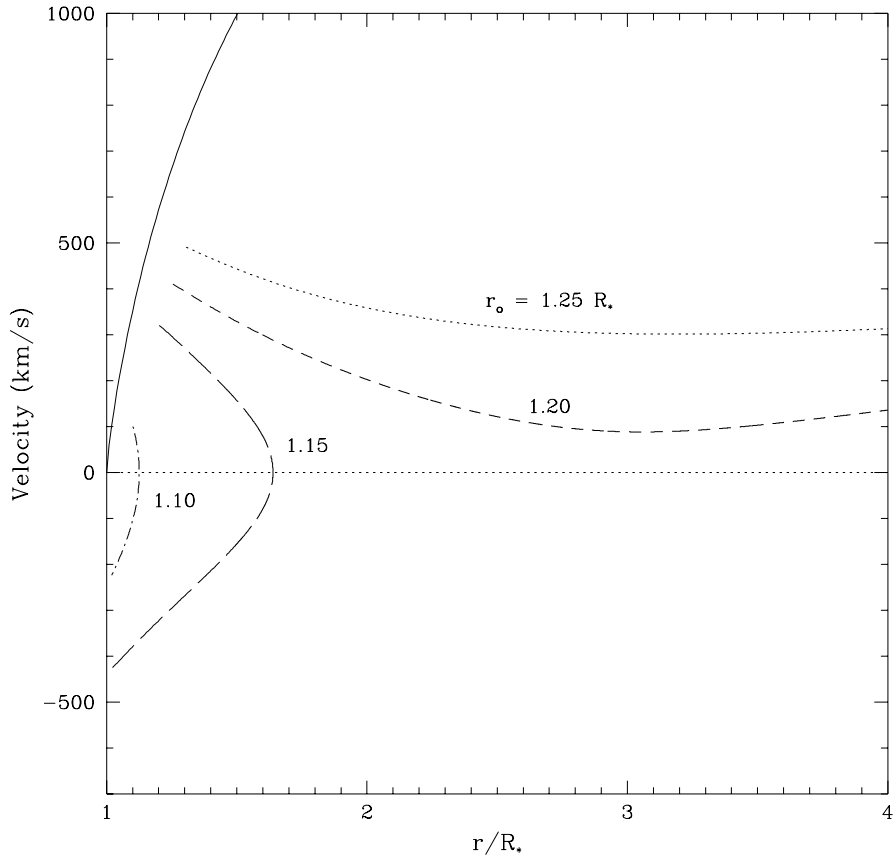}
\figcaption{As Figure \ref{fig:trajectory}, but for clumps of constant
mass and initial velocity relative to the wind [$(\log m_c, \
\omega_o) = (19.0, \ 250 \ \mathrm{km \ s^{-1}})$] with varying
initial radii [$r_o = 1.10, \ 1.20, \ 1.25, \ \mathrm{and} \ 1.30 \
R_*$].  The initial radii are labelled for each trajectory.
\label{fig:trajectory2}}
\end{figure}

\begin{figure}
\plotone{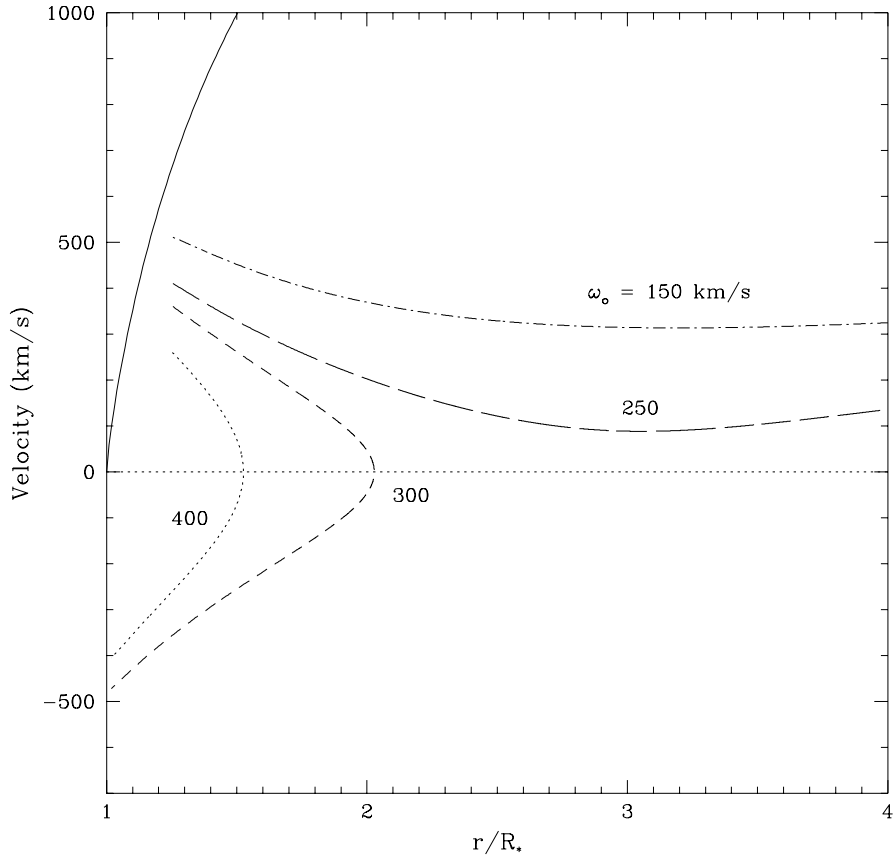}
\figcaption{As Figure \ref{fig:trajectory}, but for clumps of constant
mass and initial radii [$(\log m_c, \ r_o) = (19.0, \ 1.25 \, R_*)$]
with varying initial relative velocities [$\omega_o = 150, \ 250, \
300, \ \mathrm{and} \ 400 \ \mathrm{km \ s^{-1}})$].  At this position
in the wind, $\onehalf v_w = 331$ \kms.  The initial relative
velocities are labelled for each trajectory.
\label{fig:trajectory3}}
\end{figure}

To demonstrate the sensitivity of these trajectories to initial
conditions, we have compiled a series of trajectories that are
different in only one of the three possible input conditions.  Figure
\ref{fig:trajectory} shows the trajectories of clumps with $r_{\circ} =
1.20 \ R_*$ and $\omega_{\circ} = v_w/2 = 331$ \kms; the masses of the
clumps in this figure are $\log m_c = 18.0,$ 19.0, 19.5, and 20.0
($m_c$ in grams).  The low-mass clumps are driven outward through the
wind while the high-mass objects relatively quickly stop and start to
descend towards the star.  The low-mass clumps continue outwards
because of their lower column density (and correspondingly low
$\tau_L$), allowing for more effective radiation driving.  Lower-mass
clumps are also more effectively pushed outwards by the drag force
provided by the on-rushing stellar wind, though this is usually minor
compared with the force provided by resonance line scattering of
radiation.  Figure \ref{fig:trajectory2} shows trajectories for clumps
having the same mass and initial velocity relative to the wind [$(\log
m_c, \ \omega_o) = (19.0, \ 250 \
\mathrm{km \ s^{-1}})$] but with starting positions $r_o = 1.10$,
1.15, 1.20, and 1.25 $ R_* $; and in Figure \ref{fig:trajectory3} we
show trajectories for clumps having a mass $\log m_c = 19.0$ starting
at $r_o = 1.20 \ R_*$ with initial velocities relative to the wind of
$\omega_o = 100$, 200, 300 and 400 \kms.  Clumps with larger
$\omega_o$ tend to fall in more readily than those with low
$\omega_o$, as parcels of gas with smaller initial distances from the
stellar surface, $r_o$, are also more likely to fall inwards.  We note
that the factor $C_\rho = 0.29$ included in Equation (\ref{eqn:dens})
is not well known.  If this constant were larger, the clumps in our
models would be more effectively confined and have higher densities.
This would make it easier for a clump of given initial conditions to
fall inwards.

In Table \ref{table:traj} we give several properties for trajectories
with a range of initial conditions.  Included in the table are the
initial conditions $\omega_o$, $r_o$, and $\log m_c$.  Also given is
the lifetime ${\cal T}$ of the clump, in hours, the final velocity of
the clump in our calculations, the maximum relative velocity between
the clump and the stellar wind, $\omega_{max}$, and the corresponding
maximum shock temperature.  For the outflow trajectories, the final
velocity is the velocity of the clump at a distance of $10 R_*$ from
the surface of the star, and the lifetime is the time it takes to
reach this point.

It is clear from Figures \ref{fig:trajectory}--\ref{fig:trajectory3}
that conditions for infall can be met for a broad range of initial
conditions.  Our hypothesis that the infall of material may be
responsible for the observed red-shifted component of the
high-ionization lines is thus reasonable.  Table \ref{table:traj}
shows that many combinations of initial conditions can yield infall
trajectories. Infall velocities of $\sim300$ \kms\ (as observed in the
red wings of the \osix\ lines) are found for clumps with masses in
excess of $\sim10^{19}$ grams.  Clumps originating at larger distances
from the star reach higher infall velocities.  Clumps following infall
trajectories spend more time near their ``turning point,'' i.e., where
$v_c \simeq 0$, than close to the photosphere where their infall
velocity reaches its maximum.  The observed \osix\ profiles require
this, showing absorption consistent with a Gaussian optical depth
distribution centered at $v_c \simeq 0$ \kms\ relative to the
photosphere.

Typical lifetimes for clumps following infall trajectories are $\simeq
3-10$ hours, while outflow trajectories take $15-30$ hours to reach $r
\approx 10 R_*$.  In general clumps with initial distances $r_\circ
\ga 1.2 R_*$ with $\omega_\circ = 0.5 v_w(r_\circ)$ require masses of
order $\log m_c \ga 19.1$ to achieve infall.  The highest-temperature
gas is produced for clumps that reach the largest distances from the
stellar surface.  Thus lower mass clumps starting at relatively large
distances from the star produce the hardest X-rays.  One can see from
Table \ref{table:traj} that the trajectories followed by high-density
clumps with $\log m_c = 19.5$ tend to produce slightly harder X-rays
than those with $\log m_c = 18.0$ or 20.0.  We believe the most likely
clump masses for \tsco\ are in the range $\log m_c = 19.0$ to 19.5 (in
grams) given their ability to infall and produce very hot gas.

For infall trajectories $\omega_{max}$ occurs near the turn-around
point in a cloud's trajectory.  Specifically, the largest velocity
jumps are usually found immediately after a clump of material has
begun to fall inwards.  At this point the parcel of gas is at the
maximum height above the star, and the ambient wind velocity is larger
than at any other point in the clump's trajectory.  Though clumps may
accelerate as they near the star, the decrease in the wind velocity
with distance from the stellar surface offsets this increase in the
infall velocity of the material.

\section{X-RAY EMISSION}
\label{sec:xrays}

\subsection{Temperature Distribution}

Having shown that the interpretation of the red-shifted high-ion
absorption as infall is a physically reasonable one, we now approach
the anomolous X-ray spectrum for \tsco.  We have speculated that the
observed X-rays are a by-product of the interaction of infalling
clumps with the outgoing wind.  The temperature $T_s$ for gas in a
section of a bow shock can be calculated by substituting $v_s \cos
\theta$ for $\Delta v$ in Equation (\ref{eqn:tshock}), where $v_s$ is
the shock-jump velocity (here the relative velocity between the wind
and a clump,~$\omega$) and $\theta$ is the angle at which the material
enters the shock, relative to a ray normal to the shock's surface.
The temperature of gas is a function of position along the shock's
leading edge, via the $\theta$ dependence.  Calculating the true run
of temperatures requires knowledge of the exact shape of the
bow-shock.  This is an inherently difficult problem and is beyond the
scope of this work.  When discussing the temperatures produced by our
infalling parcels of gas, we will typically give results that assume
$\theta=0^\circ$, and, following LW, values corresponding to the shock
at $\theta = 30^\circ$. Truth may be somewhere between these values
for the majority of the shocked gas.

The trajectories calculated in \S \ref{subsec:traject} permit us to
estimate the bow-shock temperature at each point of a clump's history
given its velocity relative to the ambient wind.  It is clear that the
infall velocities of the trajectories displayed in Figures
\ref{fig:trajectory}--\ref{fig:trajectory3} can exceed $\sim 300 $
\kms, sometimes approaching $\sim 500$ \kms\ near the base of the
wind.  As mentioned in \S \ref{subsec:traject}, the relative velocity
between the wind and the infalling clumps can be much higher than
these values.  The maximum relative velocities in the trajectories for
clouds having $\log m_c = 18.0$, 18.5, 19.0, and 19.5 in Figure
\ref{fig:trajectory} are $\omega_{max} \approx 1460,$ 1660, 1295, and
1202 \kms.  These velocities imply maximum shock temperatures of order
$T_s \approx (31, \ 40, \ 24, \ \mathrm{and} \ 21) \times \e{6}$ K,
respectively, well in excess of the temperatures required to match the
CCW and Cassinelli \etal\ (1994) fits.  Even the outflow trajectories
provide very large velocity jumps, and hence large shock-jump
temperatures.  If we assume $\Delta v = \omega \cos 30^\circ$ we find
$T_s \approx (5 - 10) \times \e{6}$ K.  This is still quite close to
the temperatures required to explain the \asca\ observations of CCW.

\begin{figure}
\plotone{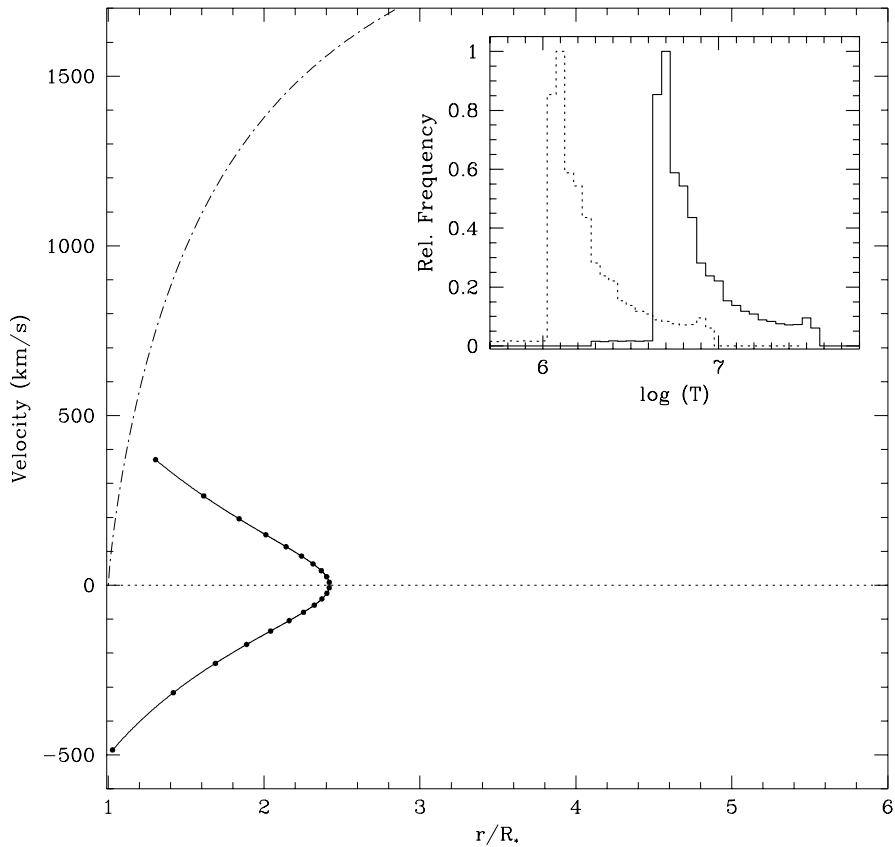}
\figcaption{The main plot shows a trajectory for a clump with the
initial conditions $(\log m_c, \ r_o, \ \omega_o) = (19.0, \ 1.30 \
R_*, \ 371 \ \mathrm{km \ s^{-1}})$.  This clump follows an infall
trajectory.  The trajectory is marked with a dot every 1.25 hours.
Also shown is the wind velocity calculated with a $\beta$ velocity law
(dot-dashed line).  The inset shows the relative distribution of
temperatures in the shock-heated gas formed as the clump follows this
trajectory.  This distribution is weighted by $\rho_w(r)^2$ for each
point along the trajectory to roughly account for the dependence of
the X-ray emission on density.  The shock-jump velocities, $\Delta v$,
are calculated from the relative velocity between wind and the clump,
$\omega$, such that $\Delta v = \omega
\cos \theta$.  The solid distribution of temperatures assumes $\theta =
0^\circ$, while the dotted distribution uses $\theta = 30^\circ$.
\label{fig:tempdist1}}
\end{figure}

\begin{figure}
\plotone{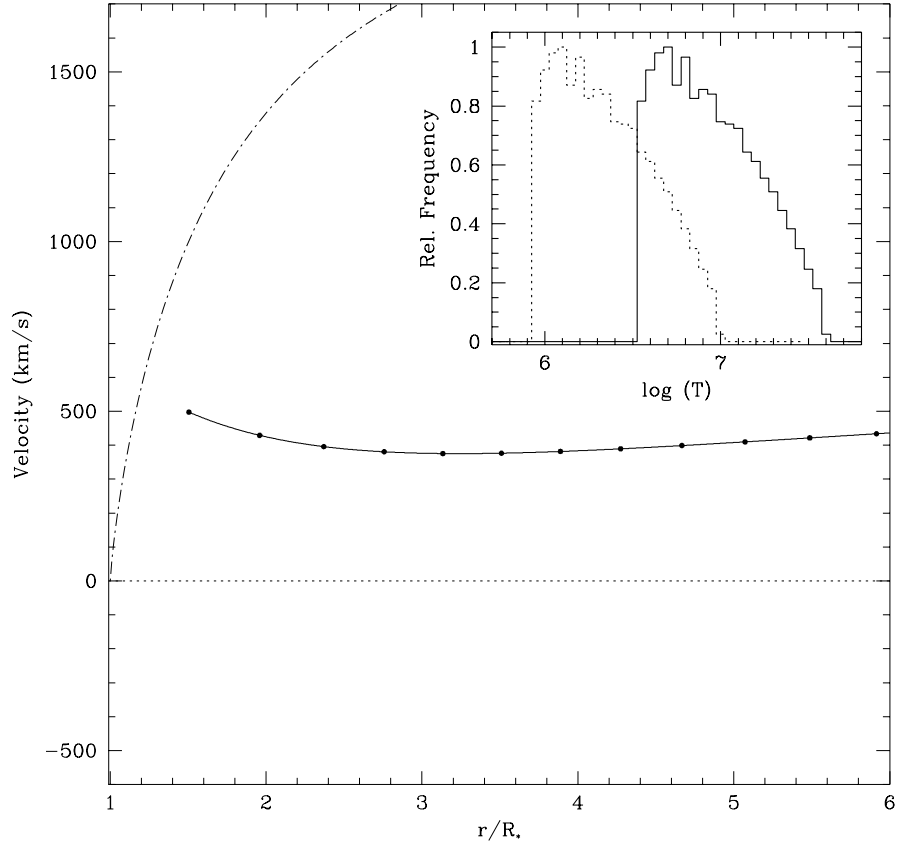}
\figcaption{As for Figure \ref{fig:tempdist1}, but for the initial
conditions $(\log m_c, \ r_o, \ \omega_o) = (19.0, \ 1.50 \ R_*, \ 498
\ \mathrm{km \ s^{-1}})$.  This clump follows an outflow trajectory.
\label{fig:tempdist2}}
\end{figure}

Figures \ref{fig:tempdist1} and \ref{fig:tempdist2} show two
trajectories as well as the model temperature distribution of the
shock-heated gas.  The initial conditions for the trajectory shown in
Figure \ref{fig:tempdist1} are $(\log m_c, \ r_o, \ \omega_o) = (19.0,
\ 1.3 \ R_*, \ 371 \ \mathrm{km \ s^{-1}})$, while those for Figure
\ref{fig:tempdist2} are $(\log m_c, \ r_o, \ \omega_o) = (19.0, \ 1.5
\ R_*, \ 498 \ \mathrm{km \ s^{-1}})$; in each case $\omega_\circ =
v_w/2$.  Changing the initial height above the stellar surface between
these two figures causes the clump to flow outwards in the wind.  The
maximum shock temperatures expected from these two trajectories are
near $5\times \e{6}$ K.  The insets show the expected {\em relative}
temperature distribution over the lifetime of the clump.  These
distributions have been weighted by $\rho_w(r)^2$ for each point in
the clumps' trajectories to approximately account for the expected
differences in the X-ray emission measure at different points in the
wind.  We show both the distribution assuming $\theta = 0^\circ$
(solid histogram) and $\theta = 30^\circ$ (dotted histogram) in these
figures.

The temperature distributions shown in Figures \ref{fig:tempdist1} and
\ref{fig:tempdist2} show that the trajectories of dense parcels of gas
derived in \S \ref{sec:infall} provide large enough shock-jump
velocities to produce gas in excess of $\e{7}$ K.  This treatment is
very approximate.  For instance we have neglected the cooling of the
gas in the bow shocks around the clumps.  The gas that is heated to $T
\sim \e{7}$ will eventually cool through the lower X-ray emitting
temperatures, providing for enhanced emission in the soft X-ray bands
over that shown in Figures \ref{fig:tempdist1} and
\ref{fig:tempdist2}.  
Though the temperature distributions peak at the same temperatures,
the temperature distribution for the infall trajectory (Figure
\ref{fig:tempdist1}) is narrower than that of the outflow trajectory 
(Figure \ref{fig:tempdist2}), which shows a strong tail towards higher
temperatures.  The ratio of the emission measures of X-ray emitting
gas observed by CCW at (1.2 and $2.4) \times \e{7}$ K is $\la 3.5$.
Both the outflow and infall trajectories have similar displayed in
these figures produce X-ray emitting material at (1.2 and $2.4)\times
\e{7}$ K in a ratio of 1.9:1.  This value is consistent with the
observational constraints, though the effects of radiative and
adiabatic cooling will affect the final distribution of X-ray
emission.

In all likelihood a population of clumps in the wind of \tsco\ will
have a range of initial conditions, providing a range of X-ray
producing temperatures and X-ray luminosities per clump.  The
distribution of intial conditions, the true shape of the bow shock at
the leading edges of the infalling clumps, and the cooling of the
post-shock gas will all determine the true temperature and luminosity
distribution of X-ray emitting gas.  The derivation of these factors
is beyond the scope of this paper.  Our goal is to show that the
production of very hot gas, in excess of $\e{7}$ K, is possible in the
wind of \tsco\ through the infall of dense material.  Our dynamical
modelling of the dense clumps in \S \ref{sec:infall} has shown that
infall can occur.  The discussion presented in this section shows that
the infall can create the requisite conditions for producing this very
hot gas

\subsection{X-ray Luminosity and the Number of Clumps}

The X-ray luminosity predicted in our model is also reasonable.  The
mechanical luminosity of the wind $L_w \sim \onehalf \mbox{\.{M}}
v_\infty^2$
%=5.5e34
is a factor of $\sim 780$ greater than $L_X$ for this star.  Even if
we assume the appropriate velocity for use is not the full terminal
velocity, but rather $\sim 1500 - 2000$ \kms, similar to the observed
wind velocities at distances typical of our model clumps, we still
have $L_w \sim 300 - 600 L_X$.  Thus the ensemble of clumps need only
convert $<1\%$ of the wind mechanical luminosity to X-ray luminosity
to match the observed $L_X$.

We estimate the available luminosity, $L_c$, for X-ray production from
a single clump as $L_c \simeq \onehalf \rho_w \omega^3 A_c$.  The
trajectory in Table \ref{table:traj} with initial conditions $(\log
m_c, \ r_o, \ \omega_o) = (19.0, 1.20 \ R_*, \ 286 \ \mathrm{km \
s^{-1}})$ reaches a maximum height of $\sim1.5 \ R_*$ with a relative
velocity $\omega \sim 1000$ \kms\ at this height.  This clump will
have an area $A_c \simeq 3\times \e{20}$ cm$^{2}$ at its maximum
height, and we find $L_c \simeq 5\times \e{29}$ ergs s$^{-1}$.  Thus
of order $\sim150$ individual clumps, assuming perfect efficiency, are
required to produce the observed X-ray luminosity of the star, $L_X
\approx 7.3 \times \e{31}$ ergs s$^{-1}$ (Cohen, Cassinelli, \&
MacFarlane 1997; hereafter CCM).  If we assume an efficiency of
$\sim10-20\%$ for conversion of kinetic energy to X-ray luminosity in
the \rosat\ bandpass (see, e.g., Wilson \& Raymond 1999), then the
total number of clumps required by the X-ray observations is of order
$\e{3}$.  The observed luminosity of the hard component is
significantly lower than the total ($L_X \sim 1.8 \times \e{31}$ ergs
s$^{-1}$; CCW), in qualitative agreement with the temperature
distributions shown in the insets of Figures \ref{fig:tempdist1} and
\ref{fig:tempdist2}.  

We can estimate a lower limit to the number of clumps from the lack of
(short-term) variability in the observed X-rays (CCM).  The population
of clumps that exist in the wind at any one time ($N$) should be large
enough to produce X-ray variations less than $\sqrt{N}/N \la 0.04$, or
$N \ga 600$ clumps.  Assuming 600 clumps of mass $\log m_c \approx
19.0$ with lifetimes similar to those in Table \ref{table:traj},
${\cal T} \sim 5-11$ hours, this yields a creation rate of mass in
clumps of $\mbox{\.{M}}^{create}_c \sim (2-5)\times \e{-9}$ \Msunyr.
Assuming half of the clumps actually escape in the wind (follow
outflow trajectories), the mass loss rate in clumps is $\mbox{\.{M}}_c
\sim (1-3)\times \e{-9} \ \mbox{\Msunyr}, or \sim 10\%$ of
\.{M}.

Another constraint on the number of clumps residing in the stellar
wind of \tsco\ comes from the \osix\ absorption line profile.  The
$\lambda1038$ \AA\ profile shows $\sim10\%$ absorption near $v\sim0$
\kms.  Assuming the \osix\ optical depth for an individual clump is much 
larger than unity and, as above, that the clumps are at a distance
$r\sim1.5 \ R_*$ with mass $\log m_c \approx 19.0$, approximately $N_c
\sim 1900$ clumps are required to cover 10\% of the total solid angle
of the star.  This corresponds to a mass loss rate of $\mbox{\.{M}} =
(0.8 - 2.0) \times \e{-8}$ \Msunyr\ for assumed lifetimes ${\cal T}
\sim 5-11$ hours.  If the efficiency of mechanical to X-ray luminosity 
from the previous paragraph is $\sim10\%$, these two estimates of the
number of clumps are in agreement.  It is worth noting that the infall
trajectories are naturally weighted to provide more absorption near
$v_c \sim 0$ \kms, i.e., near the upper-most point in the trajectory,
than at larger infall velocities.  This can be seen in Figure
\ref{fig:tempdist1} where we have plotted a dot every 1.25 hours along
the clump's trajectory.  The high density of dots near the turning
points implies that a random sample of clumps drawn from from this
trajectory will be heavily weighted towards low velocities, in
agreement with the Gaussian shape of the \osix\ absorption.  Detailed
line-profile calculations will be the subject of a future work.

Our calculations suggest that the required number of clumps in the
wind of \tsco\ is of order $\e{3}$, with \.{M}$_c$ approximately
$\mbox{a few }\times \e{-9}$ \Msunyr, consistent with the adopted mass
loss rate of this star.  These rough numbers can vary significantly
depending on the assumptions made, however.  At the present time there
is no clear evidence for an inconsistency with the observational
constraints.  Further work will need to place more stringent
constraints on the allowable parameters of this model.

\section{WHY $\tau$ SCO?}
\label{sec:conclusions}

Though our simple model can explain several of the observational
aspects of the stellar wind of \tsco, there are several questions and
details that we have not approached.  Principal among these is what
makes \tsco\ so special?  Why do most hot stars not show the observed
peculiarities of this star?  This is may be due to the special
location of \tsco\ on the main sequence where the mass loss rate
steeply drops with $T_{\rm eff}$. Main sequence stars hotter than B0,
i.e., the main sequence O stars, have mass loss rates considerably
higher than
\tsco. For instance, the mass loss rate of O9 V stars with $T_{\rm
eff} \simeq 35,000$ K, $L \simeq 10^5 L_{\odot}$, and $M
\simeq 25 M_{\odot}$ are of  order  $10^{-7}$ \Msunyr\ (predicted 
by the CAK theory), compared with less than $10^{-9}$ or $\e{-10}$
\Msunyr\ for B1.5 and later stars.  This is observed in the sudden
disappearance of the P Cygni profiles of the UV resonance lines in
stars between O9.5V and B0.5V (see the atlas of UV line profiles of
Snow et al. 1994).  The mass loss rates of B1.5 V stars are typically
a factor of $\sim 10$ below \tsco\ (e.g., the stars in CCM), with B2
star mass loss rates often being a factor of 100 lower than that
adopted for \tsco.

We speculate that the prevailing conditions for stars near the
position of \tsco\ in the HR diagram are just right not only for the
formation of dense clumps of material within the wind, but also for
their ability to infall.  Stellar winds near the spectral type of
\tsco\ are only marginally able to be driven by the radiation of the
underlying stars.  Clumps with no organized velocity gradient formed
in such winds are less likely to be pushed outward than in the winds
of earlier-type stars.

To investigate the differences between \tsco\ and earlier-type O
stars, we have calculated trajectories for high-density clumps in the
wind of the O9~V star 10 Lacertae.  We adopt stellar parameters given
in Brandt \etal\ (1998) and Snow \etal\ (1994) in our calculations:
$L_* = \e{5} \ L_\odot$, $M_* = 25$ \msun, $R_* = 8.3 \ R_\odot$,
$\mbox{\.{M}} = 1.6\times\e{-7}$ \Msunyr, $T_{eff} = 36,000$ K, and
$v_\infty \equiv 2.6 v_{esc} = 2520$ \kms.  We note that like \tsco,
10 Lac has a low projected rotation velocity ($v \sin i \approx 40$
\kms).  We find no infall trajectories for initial distances from the
stellar surface $r_o \ga 1.03 \ R_*$, initial relative velocities
$\omega_o \la 0.5 v_w(r_o)$, and masses in the range $\log m_c <
21.0$.  This is primarily a function of the increased luminosity to
mass ratio of the star, though the extra drag force plays a limited
role. This explains the lack of red-shifted absorption components of
\osix\ in the spectrum of 10 Lac.

Our calculations show that clumps in the wind of 10 Lac will have
shock jump velocities in excess of 1000 \kms, producing temperatures
in excess of $\e{7}$ K. This is very close to the value of $0.5 v_w$
found for shocks in ``normal'' stellar winds.  The emission from
(outflowing) clumps in the wind of 10 Lac will be indistinguishable
from the normal X-ray flux due to shocks in unstable stellar winds of
O-stars.

The lack of redshifted absorption profiles in the UV lines of main
sequence stars of spectral type B0.5 and later is due to drastic
decrease in mass loss. A simple scaling of the optical depth of the
\osix\ lines by a factor 0.1 compared to \tsco, due to a ten times
lower mass loss rate, makes these lines undetectable in the {\em
Copernicus} spectra, though few such stars were observed with {\em
Copernicus}.  The significantly lower mass loss rates of later-type
stars may not provide the requisite conditions for formation of an
ensemble of clumps like those thought to exist in the wind of \tsco.
In general we would expect fewer, lower mass, clumps to form in the
winds of later type stars.  Such clumps would be less dense than their
counterparts in the wind of \tsco, hence more likely to flow outward.
In the winds of later-type B stars, with significantly lower densities
and terminal velocities, the clumps would produce lower temperature
and lower luminosity X-rays, perhaps undetectable or indistinguishable
from the expected shock emission.

\section{SUMMARY}
\label{sec:summary}

We have outlined a model to explain the two peculiar observational
properties of the stellar wind of \tsco: infall of high-density clumps
surrounded by very highly-ionized gas that gives rise to the
red-shifted absorption wings of the \osix\ lines, and strong shocks
that produce the hard X-ray spectrum.

Our model hypothesizes that density enhancements in the stellar wind
lead to the formation of dense clumps of material, confined by ram
pressure, that decouple from the dynamical flow of the ambient stellar
wind.  These clumps have very large optical depths in the UV resonance
lines responsible for driving the stellar wind. The resulting
radiative acceleration of the clumps is insufficient to overcome the
gravity, so they slow down and fall back towards the star.  As they do
so, their interaction with the on-rushing stellar wind produces a bow
shock, with shock temperatures reaching several times $\e{7}$ K.

We have presented detailed dynamical models of density enhancements in
the wind of \tsco, which show that infall can be achieved for a wide
range of initial conditions.  Using these trajectories we have shown
that the shock-jump velocities can approach 2000 \kms.  The
anomolously hard X-rays observed from \tsco\ by Cassinelli \etal\
(1994) and CCW are produced in these bow shocks at temperatures of
$\sim (1-5)\times\e{7}$ K.  The infalling high-ionization material
observed in the \osix\ doublet by Lamers \& Rogerson (1978) is
material in and around the infalling clumps that is ionized by the
hard X-rays produced in the bow shocks at the clumps' leading edges.

Our calculations show that clumps with masses in the range of
$10^{19}$ to $10^{20}$ g will fall back if they are formed at radii
$\la1.3 \ R_*$ (see Table 2). Only the clumps formed above 1.1 $R_*$
produce shocks with temperatures in excess of $10^7$ K. Clumps formed
this high have maximum infall velocities of order 500 \kms. This is
higher than the observed maximum redshift of 250 \kms, but the clumps
spend very little time at such high velocities.  Even clumps with a
large final infall velocity will give the maximum contribution to the
the UV line profile at small velocities, because they spend much more
time near their turning point, where $v \simeq 0$, than at high
velocity just before they disappear into the photosphere.  This
explains why the \osix\ absorption reaches a maximum near 0 \kms.

We estimate that $\ga \e{3}$ clumps of mass $\e{19} - \e{20}$ g are
needed to explain the observed \osix\ absorption lines and the hard
X-ray flux.  This corresponds to a clump mass loss rate of
\.{M}$_{clumps} \sim 1-5 \times \e{-9} \ \mbox{\Msunyr}, or \sim 0.1 \ 
\mbox{\.{M}}_{wind}$.  In making these estimates we have assumed 
several efficiency factors whose values are not well known at present
and would require detailed modelling to be improved.  For instance,
(a) in comparing the predicted X-ray luminosities with observations we
have assumed a perfect conversion of ram pressure heating to X-ray
emission; (b) we have neglected the radiative cooling of shocks which
may result in a reduction of the predicted X-ray flux from an
individual clump; and (c) we have assumed that the clumps are
optically thick in the \osix\ transitions, and that the intrinsic line
widths of the clumps is sufficient to account for the breadth of the
observed absorption.  A reduction in these ``efficiencies'' may result
in an increase in the required number of clumps.  This will be the
subject of a subsequent study.
 
We explain the special situation of \tsco\ in terms of its location on
the main sequence where the mass loss rate drops by about factor 100
from O9 V to B1 V.  This can be tested and our model can be improved
by high S/N observations of the far UV resonance lines of \osix,
\ion{S}{6}, and \ion{P}{5} of a number of O9.5 V to B1 V stars with the 
{\em Far Ultraviolet Spectroscopic Explorer} ({\em FUSE}).  This
satellite will cover the wavelength range $905 - 1197$ \AA\ and will
observe a large number of early-type stars.  Our model would suggest
that stars with spectral types similar to \tsco\ should also show
infall in the \osix\ doublet at $\lambda \lambda 1032$ and 1038 \AA,
which will be well observed with {\em FUSE}.

The dynamical model developed in this work can be generally applied to
clumps in line-driven outflows.  Though it seems that the properties
of stars near \tsco\ on the HR diagram may be the best for producing
infalling clumps with recognizable observational consequences,
clumping may occur in many types of early-type stellar winds, and may
also be important in outflows from extragalactic sources (e.g., AGNs
and QSOs).  The dynamical model described here can be applied to all
of these cases (as discussed for the case of 10 Lac above).

\acknowledgements

We thank D. Cohen and A. Cole for many conversations regarding this
project.  JCH recognizes support from a NASA Graduate Student
Researcher Fellowship under grant number NGT-5-50121.  JPC recognizes
support from NASA under grant NAG5-2854.

%%%%%BEGIN TABLES%%%%%
\begin{deluxetable}{lcl}
\tablenum{1}
\tablecolumns{3}
\tablewidth{0pc}
\tablecaption{Stellar Parameters for $\tau$ Scorpii
	\label{table:stardata}}
\tablehead{
\colhead{Quantity}   & 
\colhead{Value}  & \colhead{Source} 
}
\startdata
Spectral Type\tablenotemark{a}      & B0.2 V     & Walborn(1994)\nl 
Distance (pc)                       & 132 & 
	{\em Hipparcos} (Perryman \etal\ 1998)  \nl 
$V$ (mag)                           & 2.82       &  
	{\em Hipparcos} (Perryman \etal\ 1998)  \nl  
$N_{\rm H I, \, ISM}$ (cm$^{-2}$)   & $2.9 \times \e{20}$ & 
	Diplas \& Savage (1994)   \nl  
\teff\ (K)                          & 31,400  & Kilian (1992)  \nl  
%
%log $g$   (cm s$^{-2}$)             & 4.24    & Kilian (1992)  \nl  
%
$R_*$     ($R_{\odot}$)             &  6.5    & Snow \etal\ (1994) \nl
$M_*$     ($M_\odot$) 	            &   15    & Snow \etal\ (1994) \nl
$v_{esc}$ (\kms)                    &  917    & Snow \etal\ (1994) \nl
$L_*$      (ergs s$^{-1}$)           &  $1.4 \times \e{38}$
					& Snow \etal\ (1994) \nl	
$L_X$\tablenotemark{b} $\ $ (ergs s$^{-1}$) 
                                    & $7.3 \times \e{31}$ 
				    & Cohen \etal\ (1997) \nl 
\vsini\ (\kms)                      & 20      & Uesugi and Fukuda (1982) \nl 

\mdot\tablenotemark{c} $\ $ (\Msunyr)   & $3.1 \times \e{-8}$ 
					& This work \nl 
$v_{\infty}$\tablenotemark{c} $\ $  (\kms) 
		                    & 2400         & This work  \nl 
\enddata
\tablenotetext{a}{The Bright Star Catalog (Hoffleit \& Jaschek 1982)
has the classification B0 V.}
\tablenotetext{b}{We have adjusted the value given in Cohen \etal\
(1997) to reflect the revised distance used here.}
\tablenotetext{c}{Theoretical calculations based on modified CAK 
theory, with values taken from   Abbott (1982) and the using the fitting
formula of   Kudritzki \etal\ (1989)} 
\end{deluxetable}

\begin{deluxetable}{cccccccc}
\tablecolumns{8}
\tablenum{2}
\tablewidth{0pc}
\tablecaption{Properties of Clump Trajectories \label{table:traj}}
\tablehead{
\colhead{$\log m_c$} & \colhead{$r_o$} & 
\colhead{$\omega_o$} & 
\colhead{$v_w(r_o)$} & 
\colhead{${\cal T}$\tablenotemark{b}} &
\colhead{$v_{b,\, final}$\tablenotemark{c}} & 
\colhead{$\omega_{max}$\tablenotemark{d}} & 
\colhead{$T_{s,\, max}$\tablenotemark{e}} \nl
\colhead{[g]} & \colhead{[$R_*$]} & \colhead{[km s$^{-1}$]} &
\colhead{[km s$^{-1}$]} & \colhead{[hrs.]} &
\colhead{[km s$^{-1}$]} & \colhead{[km s$^{-1}$]} &
\colhead{[$\e{6}$ K]}
}
\startdata
18.0 & 1.10 & 176 &  352 & \phn6.3 & $-230$ & \phn730 & \phn8 \nl
18.0 & 1.25 & 331 &  662 &    22.4 & $+790$ &    1460 & 31 \nl
19.0 & 1.10 & 176 &  352 & \phn2.8 & $-270$ & \phn580  & \phn5 \nl
19.0 & 1.15 & 235 &  470 & \phn4.5 & $-340$ & \phn820 & 10 \nl
19.0 & 1.20 & 286 &  572 & \phn6.8 & $-400$ &    1055 & 16 \nl
19.0 & 1.25 & 331 &  662 &    11.2 & $-450$ &    1295 & 24 \nl
19.0 & 1.30 & 371 &  742 &    25.9 & $-490$ &    1590 & 36 \nl
19.0 & 1.50 & 498 &  996 &    25.1 & $+540$ &    1670 & 40 \nl
19.5 & 1.10 & 176 &  352 & \phn2.5 & $-280$ & \phn560 & \phn5 \nl
19.5 & 1.15 & 235 &  470 & \phn3.8 & $-355$ & \phn785 & \phn9 \nl
19.5 & 1.20 & 286 &  572 & \phn5.5 & $-415$ &    1000 & 14 \nl 
19.5 & 1.25 & 331 &  662 & \phn7.9 & $-470$ &    1202 & 21 \nl
19.5 & 1.30 & 371 &  742 &    11.8 & $-515$ &    1405 & 28 \nl
19.5 & 1.50 & 498 &  996 &    33.7 & $+385$ &    1820 & 48 \nl
20.0 & 1.10 & 176 &  352 & \phn2.3 & $-285$ & \phn555 & \phn4 \nl
20.0 & 1.20 & 286 &  572 & \phn4.8 & $-430$ & \phn970 & 13 \nl
20.0 & 1.30 & 371 &  742 & \phn9.3 & $-530$ &    1340 & 26 \nl
20.0 & 1.50 & 498 &  996 &    58.5 & $+300$ &    1990 & 57 \nl
\enddata
\tablenotetext{a}{Velocity of the stellar wind at the initial radius,
$r_o$, of the clump.}
\tablenotetext{b}{Lifetime of the clump in hours.  For outflow
trajectories, this is the time it takes the clump to reach $r = 10
R_*$.}
\tablenotetext{c}{Final velocity of the clump.  Trajectories with
$v_{b,\, final}< 0$ are inflow trajectories, while $v_{b,\, final} >
0$ are outflow trajectories.  For outflow velocities $v_{b,\, final}$
is the velocity at $r = 10 R_*$.}
\tablenotetext{d}{Maximum relative velocity between the wind and the clump.}
\tablenotetext{e}{Maximum shock temperature obtainable given the value
of $\omega_{max}$.}
\end{deluxetable} 

%%%%%END TABLES%%%%%

\pagebreak
\clearpage

\clearpage

\end{document}